\documentclass[aps,pra,floats,floatfix,twocolumn]{revtex4}

\usepackage{ifthen}
\usepackage{ifpdf}
\usepackage{color}
\usepackage{enumerate}

\ifpdf
\usepackage{graphicx}
\usepackage{epstopdf}
\else
\usepackage{graphicx}
\usepackage{epsfig}
\fi

\usepackage{latexsym}
\usepackage{amsmath}
\usepackage{amssymb}
\usepackage{bm}
\usepackage{wasysym}

\newcommand{\mylabel}[1]{\label{#1}} 
\newcommand{\beq}{\begin{eqnarray}}
\newcommand{\eeq}{\end{eqnarray}} 
\newcommand{\be}[1]{\begin{eqnarray}\ifthenelse{#1=-1}
{\nonumber}{\ifthenelse{#1=0}{}{\mylabel{e#1}}}}
\newcommand{\ee}{\end{eqnarray}} 

\newcommand{\hide}[1]{}
\renewcommand{\cite}[1]{\textcolor{blue}{[\onlinecite{#1}}]} 

\newcommand{\ha}{\hat a}
\newcommand{\hb}{\hat b}
\newcommand{\hn}{\hat n}
\newcommand{\hu}{\hat u}
\newcommand{\hv}{\hat v}
\newcommand{\hw}{\hat w}
\newcommand{\hS}{\hat S}

\begin{document}

\title{Coherence oscillations between weakly coupled Bose-Hubbard dimers} 

\author{Christine Khripkov and Amichay Vardi}

\affiliation{Department of Chemistry, Ben-Gurion University of the Negev, Beer-Sheva 84105, Israel}

\begin{abstract}
We study theoretically the dynamics of two weakly-coupled Bose-Josephson junctions, prepared with the same particle number $N$ and Josephson excitation number $\nu$ but with different reduced one-particle purity $\gamma$. A novel entropy oscillation mode is predicted, in which one-particle coherence is transferred between the Bose-Hubbard dimers with no particle or energy transfer. We explain this purity oscillation using a semiclassical picture.

\end{abstract}

\maketitle

\section{Introduction}

The Bose-Hubbard double-dimer (BHDD), consisting of two weakly-coupled bosonic Josephson junctions, has been recently proposed as a minimal model for studying the emergence of thermodynamics in mesoscopic systems \cite{Strzys10,Strzys12a,Strzys12b,Chianca11}. The timescale separation between fast intra-dimer motion and slow inter-dimer dynamics was used to identify the internal Josephson excitations of the two dimers as a primitive form of 'heat'. This identification is reasonable because such excitations correspond to the uncontrolled shortest timescale of the system, and because of the identity in the classical limit of time and phase-space averages over equal energy subspaces. 

Microcanonical ensemble averages could thus be replaced by time averages over the fast degrees of freedom of an effective Hamiltonian adiabatic theory. It should be noted however, that the formal ergodicity of the intra-dimer motion is only incidentally obtained in this integrable model: The energy 'surfaces' of a Bose-Hubbard dimer are 1D and are thus always fully traced by classical trajectories. In this respect, a more general minimal model for heat-transfer and thermalization in mesoscopic systems is a Bose-Hubbard double-{\em trimer} whose constituent subsystems have 3D  energy surfaces, a generally mixed phase-space,  and a transition to ergodicity as each of the trimers becomes chaotic \cite{Tikhonenkov13}.

An important incentive for the study of double-dimer systems lies in recent experimental progress which enables their realization. On the one hand, few-mode Bose-Hubbard models were attained in cold atom experiments \cite{Bloch05,Gati07,Zibold10,Smerzi97}. On the other,  the same models describe the dynamics of microcavity exciton-polaritons~\cite{savona1999,carusotto13}.  Double well polariton systems \cite{wouters07,wouters08,sarchi08} were experimentally produced by using coupled micropillars \cite{abbarchi13,galbiati12} or by making use of natural well width fluctuations that lead to a weak confinement for polaritons~\cite{lagoudakis10}. Due to the existence of left and right  circular polarization modes in each well, and their coupling through an optical anisotropy  \cite{klopotowski06}, the system is described by the BHDD model~\cite{shelykh08,solnyshkov09,Khripkov14}. Both the dilute atomic quantum gas  and the exciton-polariton realizations offer fine control of the model parameters and the initial state preparation.  Efficient tracing of the four mode dynamics is enabled by atom-interferometry and ellipsometry measurements, respectively.

The collective excitations of the BHDD have been studied in Refs.~\cite{Strzys10,Strzys12a}. As illustrated below, it was found that on timescales of the order of the weak coupling between them, the two dimers can periodically exchange both particles and their elementary Josephson excitations, referred to as 'josons'.  These two slow modes were denoted as first- and second Josephson oscillations, respectively, in analogy to first- and second sound in superfluids. Moreover, because the Josephson frequency of each dimer depends on its population, first and second Josephson oscillations couple to give two composite natural modes, one of which is dominated by joson oscillations with marginal particle exchange whereas the other involves significant particle oscillations accompanied by smaller energy transfer. 

A transfer of 'heat' between subsystems is usually taken to imply a change in two quantities, energy and {\em entropy}. While introducing the concept of josons addresses the exchange of energy, it does not explicitly describe a dynamical change of entropy. Given that the low energy regime of a double-dimer system with weak coupling is integrable, we do not expect it to fully thermalize as in the chaotic double-trimer case \cite{Tikhonenkov13}. However, in addition to the particle and joson modes, there may well be another type of oscillation involving the exchange of entropy between the constituent subsystems.

In order to investigate this possibility we note that while the four mode classical motion gives rise to the oscillations of particles and excitations described above,  the full $N$-particle quantum system has many more degrees of freedom. There should consequently exist other {\em quantum} oscillation modes corresponding to the exchange of  quantities which are assumed constant in the classical description.

In this work we find the anticipated quantum mode wherein it is the {\em one-particle entropy} that oscillates back and forth between the constituent dimers. In order to excite this entropy mode, we consider a preparation in which the number of particles and excitations in the two dimers is precisely the same, but with a one-particle entropy imbalance: one dimer is internally coherent, whereas the other is spread uniformly all over the same energy contour.  Numerical solution of the quantum $N$-particle four-mode problem gives the expected coherence oscillation. The quantum results agree well with a Semiclassical picture which predicts that one-particle entropy should be exchanged between the dimers at the joson oscillation frequency.

\section{The double-dimer model} 
We consider the tight-binding Bose-Hubbard  model of two weakly coupled dimers \cite{Strzys10,Strzys12a,Strzys12b,Chianca11,shelykh08,solnyshkov09,Khripkov14},
\begin{eqnarray}
\label{Hamiltonian}
{\hat H}&=&\left(-\frac{\Omega}{2}\sum_{\sigma} \ha_{\sigma,L}^\dag \ha_{\sigma,R}
%
-\frac{\omega}{2}\sum_{\alpha} \ha_{+,\alpha}^\dag \ha_{-,\alpha}+ H.c.\right)\nonumber\\
~&~&+U\sum_{\sigma,\alpha} \hn_{\alpha,\sigma}(\hn_{\alpha,\sigma}-1)~,
\end{eqnarray}
where $\ha_{\sigma,\alpha}$ are the boson annihilation operators for the modes  $\alpha=\{L,R\}$ of dimer $\sigma=\pm$, $\hn_{\sigma,\alpha}\equiv\ha_{\sigma,\alpha}^\dag\ha_{\sigma,\alpha}$ are the mode populations, $\Omega$ and $\omega$ are,  respectively, intra-dimer and inter-dimer coupling frequencies, and $U$ is the interaction strength. We assume that $\Omega\gg\omega$ so as to obtain a timescale separation between fast internal motion and slow inter-dimer motion. 

The BHDD Hamiltonian (\ref{Hamiltonian}) can be mapped into a spin problem \cite{Tikhonenkov07} by ordering its four modes as
$$(\ha_{+,L},\ha_{+,R},\ha_{-,R},\ha_{-,L})\leftrightarrow(\ha_1,\ha_2,\ha_3,\ha_4),$$ 
and defining fifteen SU(4) generators,
\begin{eqnarray}
\hu_{j,k}&\equiv&\ha^\dag_k\ha_j+\ha_j^\dag\ha_k~,  \\
\hv_{j,k}&\equiv&(\ha^\dag_k\ha_j-\ha_j^\dag\ha_k)/i~, \qquad \qquad 1\leq k<j\leq 4 \nonumber\\
\hw_l&\equiv&\sqrt{\frac{2}{l(l+1)}}\left(\sum_{j=1}^l \hn_j - l\hn_{l+1}\right), \; 1\leq l \leq 3\nonumber
\end{eqnarray}
Constructing a pseudo-spin vector 
$$\hat{\bf S}=(\hu_{2,1}, \hu_{3,2},\hu_{4,3},\hu_{3,1},\hu_{4,2},\hu_{4,1},\hv_{2,1},\cdots,\hv_{4,1},\hw_1,\hw_2,\hw_3)$$
 from these generators, the BHDD Hamiltonian (\ref{Hamiltonian}) assumes the form,
\begin{equation}
{\hat H}=-\frac{\Omega}{2}(\hS_1+\hS_3)-\frac{\omega}{2}(\hS_2+\hS_6)+\frac{U}{2} (\hS_{13}^2+\hS_{14}^2+\hS_{15}^2)~,
\label{HamSpin}
\end{equation}
where we have eliminated an insignificant $UN(N/4-1)$ $c$-number shift. The Heisenberg equations of motion for the $S_i$ operators thus read,
\begin{eqnarray}
\label{eom}
i{\dot \hS}_i&=&-\frac{\Omega}{2}\sum_{j=1,3}\sum_{k=1}^{15} c_{ij}^k \hS_k -\frac{\omega}{2}\sum_{j=2,6}\sum_
{k=1}^{15}
 c_{ij}^k \hS_k \nonumber\\
~&~&+ \frac{U}{2}\sum_{j=13}^{15}\sum_
{k=1}^{15} 
c_{ij}^k \Big(\hS_j \hS_k+\hS_k \hS_j\Big),
\end{eqnarray} 
where $c_{ij}^k$ are SU(4) structure constants. We define two characteristic dimensionless interaction parameters $u_\omega=UN/\omega$ and $u_\Omega=UN/\Omega$ quantifying the effect of nonlinearity/interactions on the inter-dimer and intra-dimer dynamics, respectively.

Following the weakly-coupled subsystems paradigm, we extract reduced SU(2) representations for the two $\sigma=\pm$ Bose-Hubbard dimers.  The corresponding SU(2) pseudospin vectors are,
\begin{eqnarray}
{\hat{\bf S}}_+&=& (\hS_1,\hS_7,\hS_{13})~, \\
{\hat{\bf S}}_- &=&(\hS_3,-\hS_9,\sqrt{1/3}\hS_{14}-\sqrt{2/3}\hS_{15})~,
\end{eqnarray}
 whereas the number of particles in each dimer is 
$$\langle{\hat N}_{+(-)}\rangle=\langle{\hat n}_{1(4)}\rangle+\langle{\hat n}_{2(3)}\rangle~.$$

We thus obtain two Bloch vectors ${\bf s}_\sigma=\langle\hat{\bf S}_\sigma\rangle/ N_\sigma$ where $s_{\sigma,z}$ corresponds to the normalized population imbalance within dimer $\sigma$, and $s_{\sigma,x},s_{\sigma,y}$ are respectively, the real and imaginary parts of the coherence between the dimer modes.\\

\begin{figure}[t]
\includegraphics[width=\hsize]{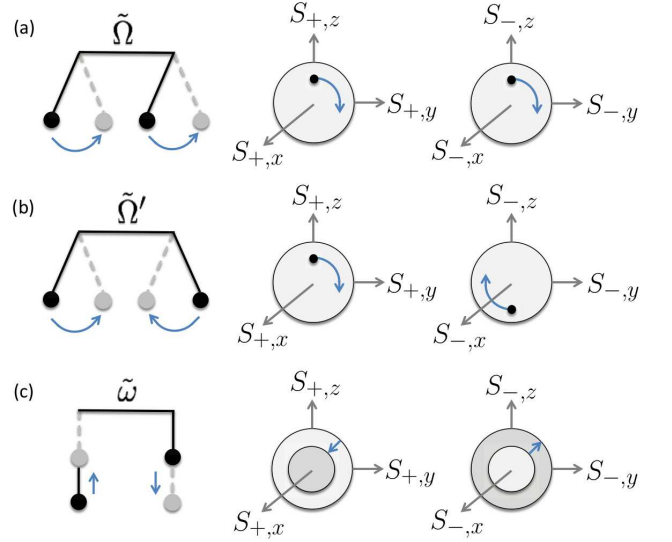}\\
\caption{Bogoliubov modes of the double dimer system: (a) fast antisymmetric oscillation, (b) fast symmetric oscillation, and (c) slow particle exchange. Left panels illustrate the equivalent coupled nonrigid pendulums motion whereas the middle and right panels depict the motion of the two reduced dimer Bloch vectors ${\bf S}_+$ and ${\bf S}_-$ on the surface of their respective Bloch spheres.}
\label{f1}
\end{figure}

\section{Coherent States and the Classical Mean-Field Limit}
The coherent states of the BHDD dimer are given as,
\begin{equation}
\left |\alpha_1,\alpha_2,\alpha_3,\alpha_4\right \rangle=\frac{1}{N!}\left[\sum_{i=1}^4 \alpha_i \ha_i^\dag\right ]^N \left | {\rm vac}\right\rangle~,
\end{equation}
where $\alpha_i=\sqrt{n_i/N}\exp(-i\varphi_i)$ are $c$-number coefficients and $|{\rm vac}\rangle$ is the vacuum state. Using the $U(1)$ gauge invariance and total number conservation, 
\begin{equation}
\sum_{i=1}^4 \left | \alpha_i \right |^2 = 1~,
\end{equation}
these states depend only on three relative populations and three relative phases, e.g., 
\begin{widetext}
\begin{eqnarray}
\label{cspar2}
\left | \Delta n_+,\Delta n_-, \Delta N, \varphi_+, \varphi_-, \Phi \right\rangle&=%
&\left\{\left[\sqrt{\frac{N+\Delta N+2\Delta n_+}{4N}}\ha_{1}^\dag +%
 \sqrt{\frac{N+\Delta N-2\Delta n_+}{4N}}\exp(i\varphi_+)\ha_{2}^\dag\right]\right. \\
~&~&\left. +\exp(i\Phi)\left[ \sqrt{\frac{N-\Delta N+2\Delta n_-}{4N}}\ha_{4}^\dag +%
 \sqrt{\frac{N-\Delta N-2\Delta n_-}{4N}}\exp(i\varphi_-)\ha_{3}^\dag  \right]\right\}^N  \left | {\rm vac}\right\rangle~,\nonumber
\end{eqnarray}
\end{widetext}
where $\mbox{$\Delta n_{+(-)}={n}_{1(4)}-{n}_{2(3)}$}$ and $\mbox{$\varphi_{+(-)}=\varphi_{2(3)}-\varphi_{1(4)}$}$ are, respectively, the internal L-R population imbalance and the relative phase within dimer +(-). The interdimer population imbalance and relative phase are $\mbox{$\Delta N=N_+ -N_-$}$ and $\mbox{$\Phi=\varphi_4-\varphi_1$}$, respectively.

To the extent that the BHDD dynamics can be restricted to the coherent states (\ref{cspar}), we can replace the operators $\ha_i$ or $\hS_j$ by the corresponding $c$-numbers $\alpha_i$ or $S_j$, to obtain the classical (mean-field) limit of the $N$ particle quantum dynamics given in Eq.~(\ref{eom}). In the following section we review the linearization of this dynamics about the coherent ground state to give the natural oscillation modes, as outlined in Refs.~\cite{Strzys10,Strzys12a,Strzys12b}. Then, in Section V, we propose an incoherent, quantum preparation which leads to coherence oscillations beyond the classical limit. 

\section{Classical Josephson modes} 
The energy spectrum of a two-mode Bose-Hubbard dimer is described in Refs.~\cite{Boukobza09,Chuchem10,Khripkov13} for various interaction regimes. Accordingly, the lower energy levels for each of the two weakly-coupled Bose-Hubbard dimers in the Rabi regime $u_\Omega<1$ and in the Josephson regime $1<u_\Omega< N^2$ \cite{Gati07,Zibold10,Smerzi97}, are harmonic-oscillator-like with Josephson frequency spacing,
\begin{equation}
{\tilde\Omega}_\sigma=\sqrt{\Omega(\Omega+2UN_\sigma)}~.
\end{equation}  
The corresponding low-energy classical trajectories are small Josephson oscillations about the nearly coherent ground state, i.e., circles about the $S_x$ axis of each dimer. In the extreme strong interaction Fock regime $u_\Omega> N^2$ the system undergoes a Mott transition, the ground state itself becomes a Fock state (corresponding to a circle about $S_z$) and the entire dimer spectrum is nonlinear. In what follows we will assume that $u_\Omega < N^2$ so that the low energy internal motion of each dimer  is simple Josephson oscillations.

The BHDD system in this linear regime, is thus equivalent to two coupled non-rigid pendulums which can exchange their length, corresponding to the {\em number of particles}  $N_\sigma$ in each dimer, as well as their oscillation amplitude, corresponding to the excitation of each oscillator. The degree of this Josephson excitation is quantified by the respective {\em number of josons},
\begin{equation}
\nu_\sigma = \frac{E_\sigma - E_{\sigma,g}}{\tilde\Omega_\sigma}= \frac{1}{\tilde\Omega_\sigma} \left[-\frac{\Omega}{2} \langle \hS_{\sigma,x}\rangle+\frac{U}{2} \langle\hS_{\sigma,z}^2\rangle+\frac{\Omega N_\sigma}{2}\right]
\end{equation}
where $E_\sigma$ and $E_{\sigma,g}=-\Omega N_\sigma /2$ denote, respectively, the energy and ground-state energy of dimer $\sigma$. While the total number of josons $\nu=\nu_+ + \nu_-$ is not strictly conserved, conservation of energy implies it is an adiabatic invariant in the linear regime.

\begin{figure}[t]
\includegraphics[width=\hsize]{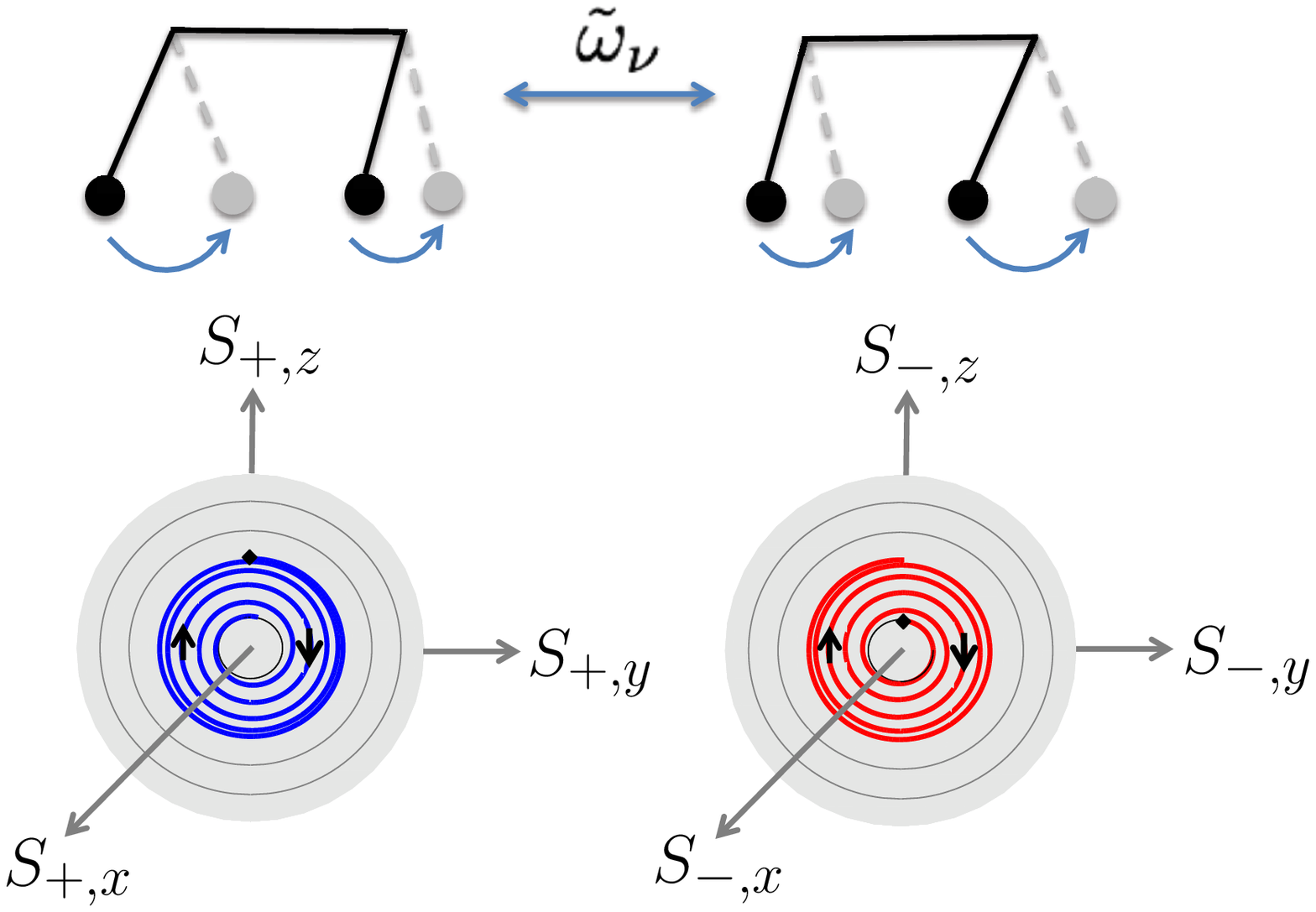}\\
\caption{(Color online) Beating of the two fast Bogoliubov modes results in a slow adiabatic joson exchange mode, corresponding to the transfer of excitations between dimers.}
\label{f2}
\end{figure}

Linearizing the BHDD Hamiltonian (\ref{HamSpin}) about its ground state, one obtains three Bogoliubov collective-oscillation modes \cite{Strzys10,Strzys12a,Strzys12b}, schematically illustrated in Fig.~\ref{f1}:
\begin{enumerate}[(a)]
\item
An antisymmetric mode where the  Josephson oscillations  of the two dimers are in-phase, at the single-pendulum frequency,
\begin{equation}
{\tilde\Omega}=\sqrt{\Omega(\Omega+UN)}~.
\end{equation}

\item
A symmetric mode with Josephson oscillations of opposite phase in the two dimers, at frequency 
\begin{equation}
{\tilde\Omega}'=\sqrt{(\Omega+\omega)(\Omega+\omega+UN)}~.
\end{equation}

\item
Slow particle-oscillations between the two polarizations, maintaining a fixed total population $N$, with  frequency 
\begin{equation}
{\tilde\omega}=\sqrt{\omega(\omega+UN)}~.
\label{particlefrequency}
\end{equation}

\end{enumerate}

In the adiabatic theory formulated in Refs. ~\cite{Strzys10,Strzys12a,Strzys12b}, the beating of the two fast modes (a) and (b) at the difference frequency, 
\begin{equation}
\omega_\nu={\tilde\Omega}'-\tilde{\Omega}\approx(\omega/{\tilde\Omega})(\Omega+UN/2)~,
\label{beatfrequency}
\end{equation}
serves as an effective second slow mode in which energy (i.e., josons) is exchanged between the dimers (see Fig.~\ref{f2}).  The frequency of this beat mode is shifted as, 
\begin{equation}
{\tilde\omega}_\nu=\sqrt{\omega_\nu(\omega_\nu+U_\nu \nu)}~,
\label{josonfrequency}
\end{equation}
by the effective attractive interaction between josons \cite{Strzys10},
\begin{equation}
U_\nu=-(U/4)(4\Omega+UN)/(\Omega+UN)~.
\end{equation}

\begin{figure}[t]
\includegraphics[width=\hsize]{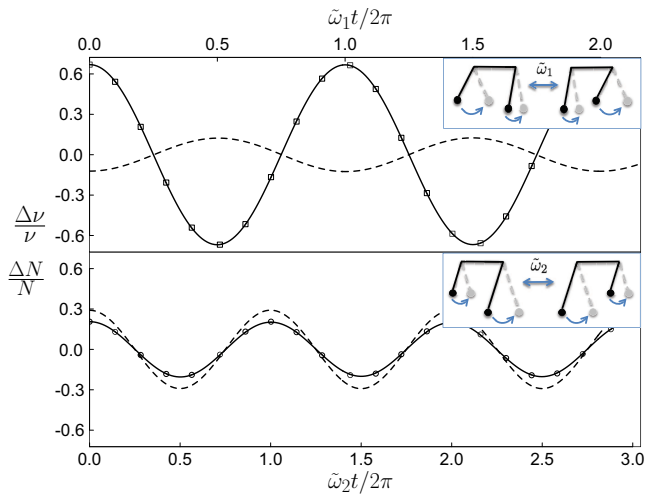}\\
\caption{(Color online) Natural  modes obtained from diagonalization of the coupled particle and joson slow modes. Dashed lines depict the particle imbalance between the dimers, whereas solid lines correspond to the joson/excitation imbalance. Markers denote oscillation at frequencies $\tilde\omega_1$ ($\square$)  and $\tilde\omega_2$ ($\circ$). Insets illustrate the equivalent exchange of length and oscillation amplitude between coupled pendulums.}
\label{f3}
\end{figure}

The two slow modes, namely the particle oscillation mode of Eq.~(\ref{particlefrequency}) and the joson oscillation mode of Eq.~(\ref{josonfrequency}), are further coupled because the frequency of oscillation within each dimer depends on its population. This leads to an effective particle-joson coupling with strength, 
\begin{equation}
U_c=(U/2)(\Omega/{\tilde\Omega})\sqrt{N\nu}~.
\end{equation}
Thus, the natural classical modes of the BHDD are obtained by diagonalization of the effective adiabatic Hamiltonian for coupled particle- and joson oscillators, resulting in two natural frequencies  \cite{Strzys10,Strzys12a,Strzys12b},
\begin{equation}
\tilde{\omega}_{1,2}^2=\frac{{\tilde\omega}^2+{\tilde\omega}_\nu^2}{2}\mp\left [ \left( \frac{{\tilde\omega}^2-{\tilde\omega}_\nu^2}{2}\right)^2+\frac{\omega\omega_
\nu\Omega U^2 N\nu}{\Omega+UN}\right]^{1/2}~.
\label{naturalfrequencies}
\end{equation}

In Fig.~\ref{f3}, we plot the classical evolution of the particle imbalance,
\begin{equation}
\Delta N\equiv N_+ - N_-
\end{equation}

\begin{figure}[t]
\includegraphics[width=\hsize]{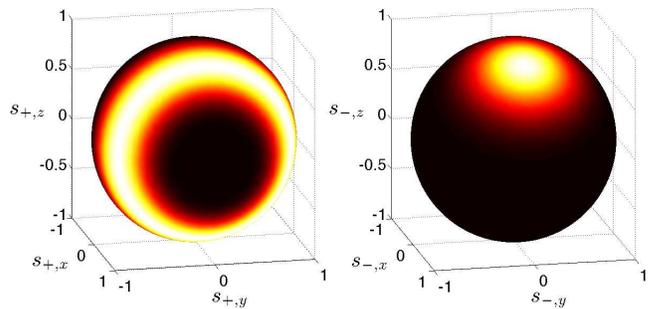}\\
\caption{(Color online) Husimi phase-space distribution for the initial quantum preparation $\left | \Psi_0 \right\rangle$ with vanishing particle-imbalance and joson-imbalance, $\Delta N=\Delta \nu=0$ but with a one-particle entropy gradient $\Delta\gamma_2$ between the two dimers. Dimer '$-$' is in a coherent state whereas dimer '$+$' is in an odd-even Fock state with the same energy.}
\label{f4}
\end{figure}

\noindent and the joson imbalance,
\begin{equation}
\Delta\nu\equiv\nu_+ - \nu_-
\end{equation}
between the dimers, for initial conditions designed to excite the ${\tilde\omega}_1$ mode (top) and ${\tilde\omega}_2$ mode (bottom). Due to the particle-joson coupling, the two natural modes involve both particle and joson oscillations.  The slower oscillation at frequency ${\tilde\omega}_1$, involves predominantly energy-exchange,  accompanied by small population oscillations. The particle- and joson oscillations in this mode are of opposite phase, i.e., the joson imbalance  is maximized when the particle imbalance is at minimum. By contrast the faster mode at frequency  ${\tilde\omega}_2$, has significant particle oscillations. In this case, particles and josons oscillate in-phase. As expected, in the zero interaction limit $U\rightarrow 0$,  the particle and joson oscillation separate and we have $\omega_1\rightarrow{\tilde\omega}_\nu$ and $\omega_2\rightarrow{\tilde\omega}$.

\section{Entropy Gradient Preparation}

In this work we consider a highly non-classical preparation in which both dimers have the same number of particles (i.e., $\Delta N=0$) and the same excitation energy (i.e., identical number of josons $\Delta\nu=0$), but differ in their {\em one-particle entropy}. In terms of their phase-space representations, one dimer resembles a coherent minimal Gaussian localized around a point on the $\nu_-=\nu/2$ energy contour, whereas the other is smeared uniformly over the $\nu_+=\nu/2$ energy contour. To construct such states, we first represent the BHDD coherent states in the even-odd ($x$) Fock basis, as, 

\begin{widetext}
\begin{eqnarray}
\label{cspar}
\left | \Delta {\tilde n}_+,\Delta {\tilde n_-}, \Delta N, {\tilde\varphi}_+, {\tilde\varphi}_-, \tilde\Phi \right\rangle&=%
&\left\{\left[\sqrt{\frac{N+\Delta N+2\Delta {\tilde n}_+}{4N}}\hb_{1}^\dag + \sqrt{\frac{N+\Delta N-2\Delta {\tilde n}_+}{4N}}\exp(i{\tilde \varphi}_+)\hb_{2}^\dag\right]\right. \\
~&~&\left. +\exp(i\tilde\Phi)\left[ \sqrt{\frac{N-\Delta N+2\Delta {\tilde n}_-}{4N}}\hb_{4}^\dag + \sqrt{\frac{N-\Delta N-2\Delta {\tilde n}_-}{4N}}\exp(i{\tilde \varphi}_-)\hb_{3}^\dag  \right]\right\}^N  \left | {\rm vac}\right\rangle~,\nonumber
\end{eqnarray}
\end{widetext}
where 
$$\hb_{1(4)}= \frac{\ha_{1(4)}+ \ha_{2(3)}}{\sqrt{2}}$$
and
$$\hb_{2(3)}= \frac{\ha_{1(4)}- \ha_{2(3)}}{\sqrt{2}}$$
are boson annihilation operators for the even and odd superpositions, respectively, of the two modes constituting dimer $+(-)$. The internal number differences are $\Delta{\tilde n}_+(-)={\tilde n}_{1(4)}-{\tilde n}_{2(3)}$ with ${\tilde n}_i=\langle\hb_i^\dag\hb_i\rangle$, whereas the corresponding relative even-odd phases within dimer $+(-)$ are ${\tilde\varphi}_{+(-)}$. We then integrate over all values of  the internal odd-even phase of dimer $+$ :
\begin{equation}
\left | \Psi_0\right \rangle = \int d{\tilde\varphi}_+ \exp(-i {\tilde n}_2 {\tilde\varphi}_+)
\left | \Delta {\tilde n}_+,\Delta {\tilde n_-}, \Delta N, {\tilde\varphi}_+, {\tilde\varphi}_-, \tilde\Phi \right\rangle~,
\label{init} 
\end{equation}
to obtain a state which has well-defined interdimer phase $\tilde\Phi$ and internal odd-even ${\tilde \varphi}_-$ phase in one  dimer, but has no internal odd-even coherence in the other, as the reduced '$+$'-dimer state is an odd-even Fock state (an eigenstate of $\hS_{+,x}=\hS_1$).  Below we set $\Delta N=0$, $\tilde\Phi={\tilde\varphi}_-=0$, and $\Delta{\tilde n}_+=\Delta{\tilde n}_-=\nu/2$. The phase-space Husimi distribution for the preparation (\ref{init}) is plotted in Fig.~\ref{f4}. In the $\sigma=-$ dimer the distribution resembles a minimal gaussian indicating perfect coherence. By contrast, in the $\sigma=+$ dimer the distribution is smeared evenly on  a ring about the $S_{+,x}$ axis, corresponding to an odd-even number state. In the $u_\Omega\ll 1$ regime, such a ring is an excellent approximation to an energy contour with a well-defined joson-number (Note that $u_\Omega\ll 1$  does not necessarily imply negligible interaction/nonlinear effects since $u_\omega$ can still be quite large as in the numerical calculations below).

\begin{figure}[t]
\includegraphics[width=0.95\hsize]{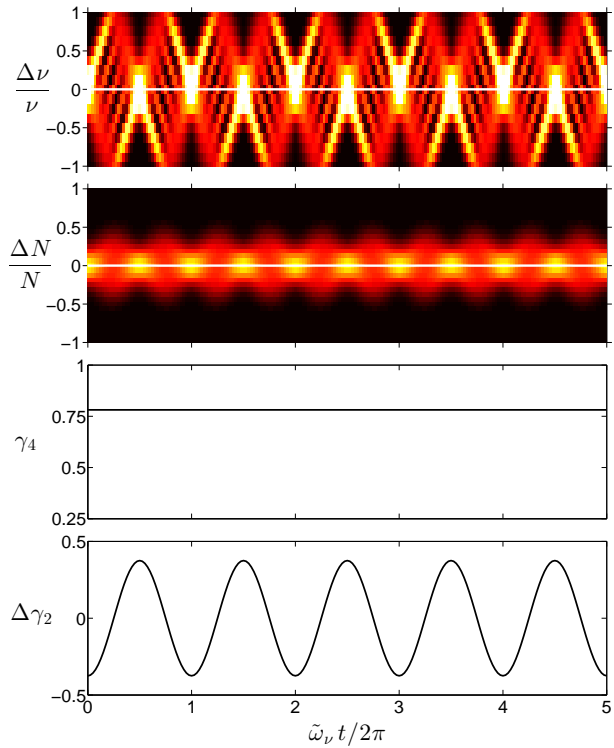}\\
\caption{(Color online) Full quantum dynamics of (from top) the joson imbalance $\Delta\nu$, the particle imbalalnce $\Delta N$, the four-mode  one-particle purity $\gamma_4$ and the interdimer purity imbalance $\Delta\gamma_2$. Solid lines in top two panels correspond to the average joson-difference and particle-difference, respectively. Parameters are $N=32$, $\nu=8$, $\Omega=10\omega$ and $UN=0$ }
\label{f5}
\end{figure}

\begin{figure}[t]
\includegraphics[width=0.95\hsize]{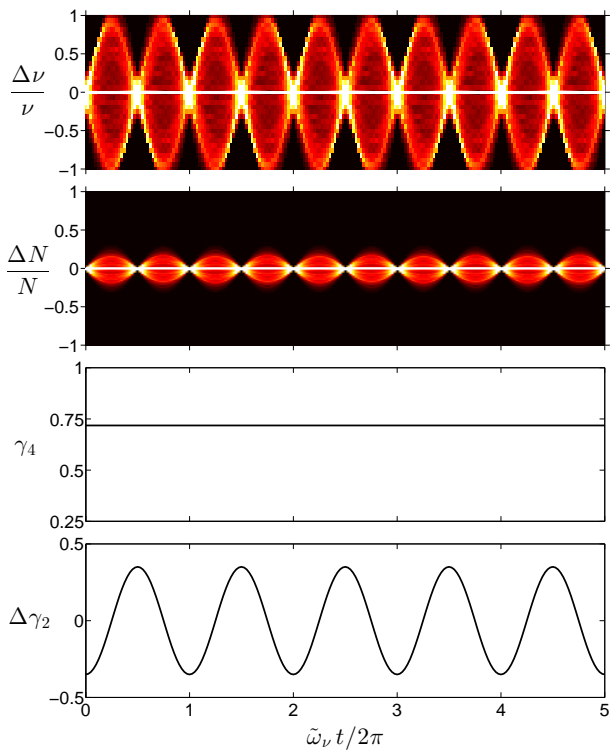}\\
\caption{(Color online) Semiclassical dynamics for the same parameters as in Fig.~\ref{f5}. }
\label{f6}
\end{figure}

\begin{figure}[t]
\includegraphics[width=0.95\hsize]{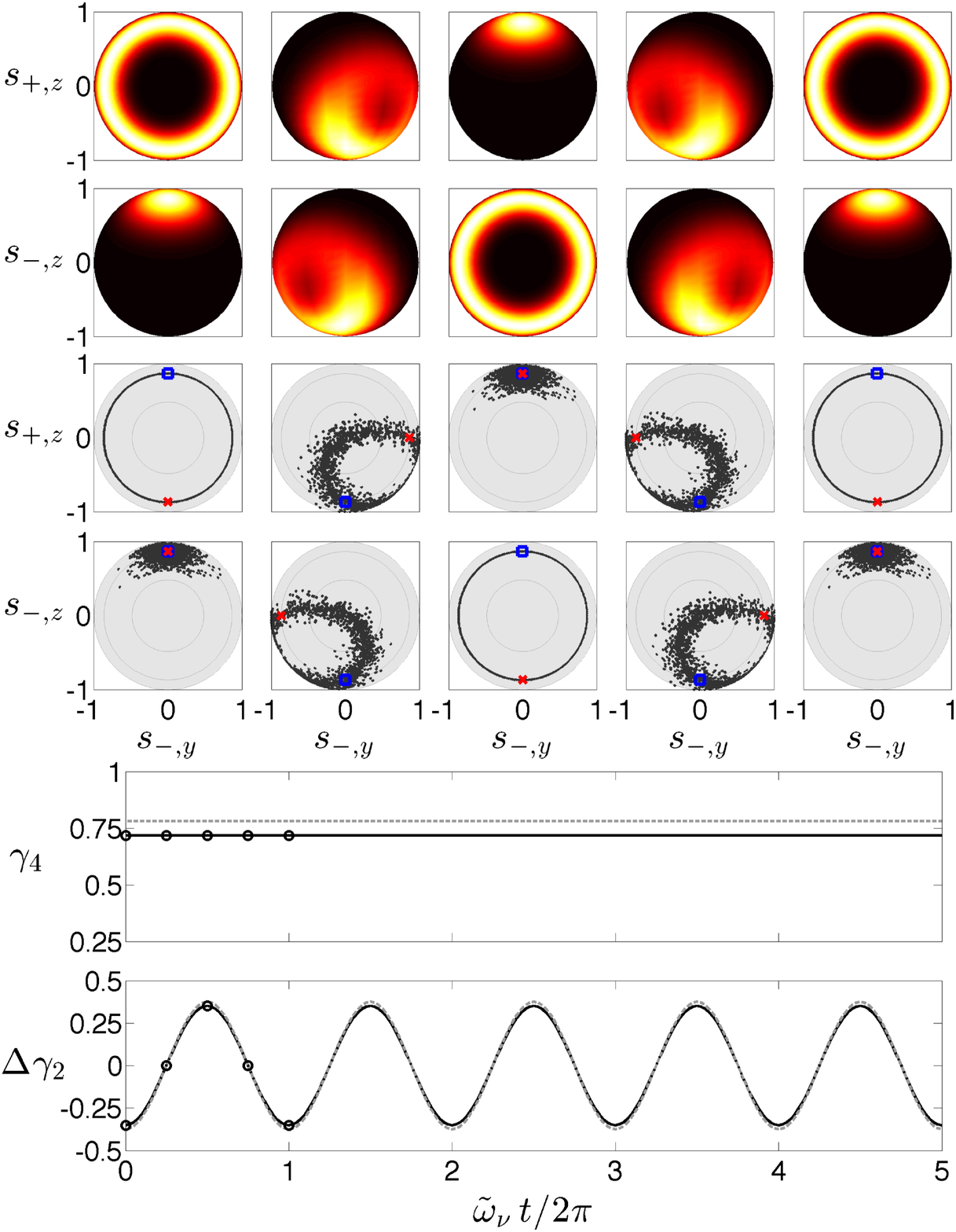}\\
\caption{(Color online) Evolution of the quantum (top) and semiclassical (middle) phase space distributions of the two coupled dimers. The resulting overall purity and interdimer purity imbalance are compared below (the dashed gray line corresponds to the quantum simulation whereas the solid black line is obtained from the semiclassical propagation). Squares in the semiclassical distribution plots denote classical points oscillating at frequency $\Omega$ whereas crosses denote classical points oscillating at frequency $\Omega'$. Circles in lower panels denote the times at which the phase-space distributions of the top panels are plotted, in order of appearance.}
\label{f7}
\end{figure}

\begin{figure}[!tb]
\includegraphics[width=0.95\hsize]{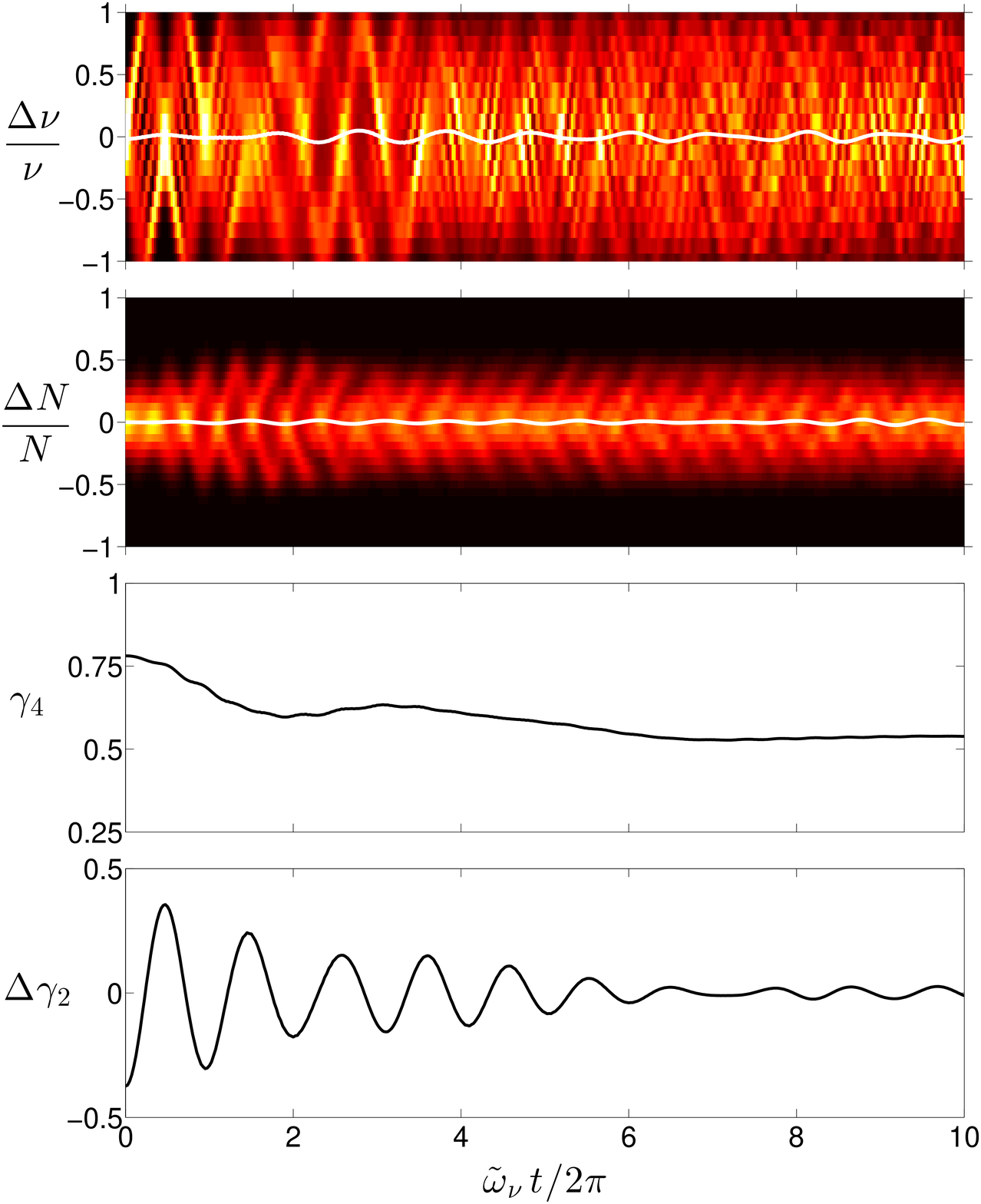}\\
\caption{(Color online) Same as Fig.~\ref{f5} with $u_\omega=0.5$.}
\label{f8}
\end{figure}

\section{Coherence Oscillations}

We numerically propagate the initial state $|\Psi_0\rangle$ according to the full quantum $N$-particle equations of motion (\ref{eom}). The preparation (\ref{init}) is approximated by a normalized superposition of $M=20$ coherent states with $\Delta N=0$, $\tilde\Phi={\tilde\varphi}_-=0$, $\Delta{\tilde n}_+=\Delta{\tilde n}_-=\nu/2$, evenly spread at ${\tilde\varphi}_+= 2\pi j/M,~j=1,\dots,M$ and multiplied by the appropriate $\exp(-i{\tilde n}_2 {\tilde\varphi}_+)$ phase factor. In Fig.~\ref{f5} we plot the linear, interaction-free evolution of the distributions $P(\Delta N)$ (probability to find a particle imbalance $\Delta N$ between the dimers) and $P(\Delta\nu)$ (probability to find a joson/excitation imbalance $\Delta\nu$ between the dimers), as well as the average particle and joson difference. The one-particle coherence of the entire system is quantified by the four-mode purity,
\begin{equation}
\gamma_4= {\rm Tr} \left\{ {[\rho^{(sp)}]^2} \right\} = \frac{1}{4} \left( 1+ 3|\bf{s}|^2\right)~,
\end{equation}
where $\rho^{(sp)}$ is the four-mode reduced one-particle density matrix and ${\bf s}=\langle {\bf S}\rangle/\sqrt{3N/2}$ is the normalized four-mode Bloch vector. Similarly, the one-particle coherence in each dimer is characterized by its two-mode purity,
\begin{equation}
\gamma_{2,\sigma}= {\rm Tr} \left\{[\rho_\sigma^{(sp)}]^2 \right\} = \frac{1}{2} \left( 1+ |\bf{s}_\sigma|^2\right)~,
\end{equation}
where $\rho_\sigma^{(sp)}$ is the reduced two-mode one-particle density matrix of dimer $\sigma$ and ${\bf s}_\sigma$ is the respective one-dimer Bloch vector defined in Section~II. Purity values of $\gamma_j=1$ correspond to full one-particle $j$-mode coherence whereas values of $\gamma_j=1/j$ indicate no such coherence. The lowest panel of Fig.~\ref{f5} depicts the time evolution of the interdimer coherence imbalance $\Delta\gamma_2=\gamma_{2,+} - \gamma_{2,-} $.

It is clear from Fig.~\ref{f5} that the average particle and excitation difference remain zero throughout the evolution. The global one-particle coherence is also preserved because there are no interactions to entangle or disentangle particles. However, the internal coherence within each dimer is not conserved and the two dimers periodically exchange their one-particle purity at the beat frequency ${\tilde\omega}_\nu$ (which in the interaction-free case is trivially equal to $\omega$). Similar results are obtained from semiclassical simulations (see Fig.~\ref{f6}), where we propagate an ensemble of classical (i.e., mean-field) trajectories, with initial-conditions distributed according to the phase-space density of the quantum preparation (\ref{init}).

\begin{figure}[!tb]
\includegraphics[width=0.95\hsize]{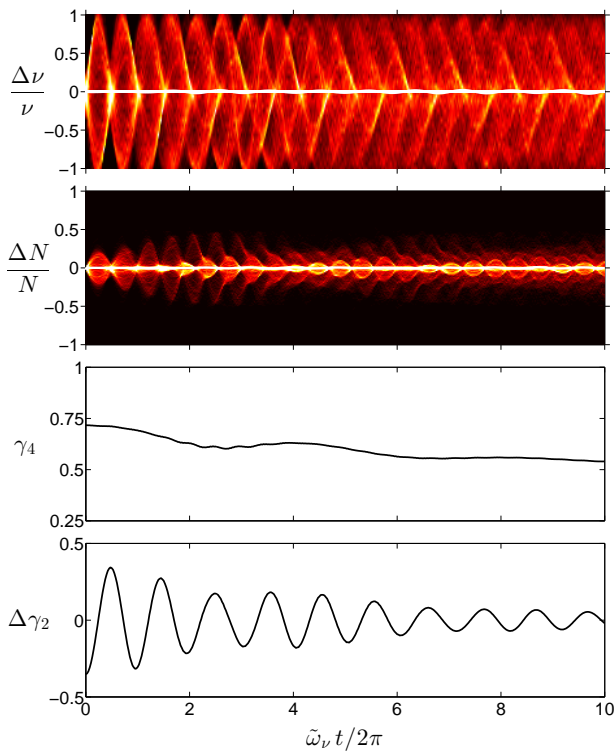}\\
\caption{(Color online) Same as Fig.~\ref{f6} with $u_\omega=0.5$. }
\label{f9}
\end{figure}

The origin of the observed coherence oscillations becomes clear when we inspect the time evolution of the quantum (Husimi) and semiclassical (Liouville) phase-space distributions (see Fig.~\ref{f7}). The square markers in the semiclassical distribution correspond to classical initial conditions which purely excite the fast antisymmetric oscillation (see Fig.~\ref{f1}a). Similarly, cross  markers denote classical initial conditions which excite pure symmetric oscillation (Fig.~\ref{f1}b). Due to the frequency difference $\Omega' - \Omega=\omega_\nu$ between these two modes, points starting together in the $\sigma=-$ dimer will reach opposite positions on the excitation-energy contour at $t=\pi/\omega_\nu$, whereas the two markers in the $\sigma=+$ dimer will coincide at the same time. Thus, the coherent distribution in the $\sigma=-$ dimer smears uniformly around the energy contour, while the smeared $s_{+,z}$ number-state distribution in the $\sigma=+$ dimer localizes to give a coherent state at the same time.

\begin{figure}[!tb]
\includegraphics[width=0.95\hsize]{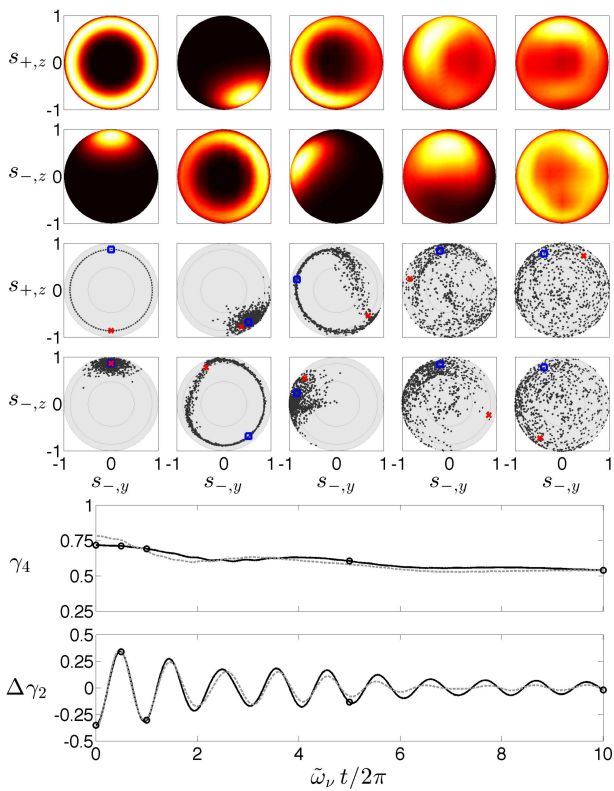}\\
\caption{(Color online) Same as Fig.~\ref{f7} with $u_\omega=0.5$.}
\label{f10}
\end{figure}

In order to see whether the one-particle coherence oscillation mode transforms in the presence of interactions in the same way as do the Josephson modes of Section IV, we repeat the calculation with $u_\omega=0.5$ and $u_\Omega=0.05$. The interaction strength was chosen so that linearization can still be expected to be valid, yet with substantial renormalization of the slow oscillation beat frequency from $\omega_\nu$ of Eq.~(\ref{beatfrequency}) to the joson frequency $\tilde\omega_\nu$ of Eq.~(\ref{josonfrequency}). The results are presented in Figs.~\ref{f8}-\ref{f10}. It is evident that unlike the linear case, the overall four-mode purity is not conserved, because different points incur different mean-field shifts which eventually lead to the spreading of the distribution throughout phase-space. However, within this decoherence time we obtain coherence oscillations between the dimers at frequency $\tilde\omega_\nu$ in both the quantum and the semiclassical propagation. There is generally a remarkable agreement between the semiclassical and quantum phase-space distributions of Fig.~\ref{f10}.

The predicted coherence oscillations can readily be detected experimentally via atom interferometry within the two dimers. The one-particle purity is a measure of the fringe visibility expected in such experiments. If the BHDD system is prepared as described in Section~V and the modes of either dimer are allowed to interfere, we expect a temporal oscillation in the observed fringe visibility with frequency $\tilde\omega_\nu$. 

\section{Experimental Realization}

There are several options to realize double-dimer systems. The exciton-polariton realization is discussed in detail in Refs.~\cite{shelykh08,solnyshkov09,Khripkov14}. With cold atoms, one possibility is to construct a four-well system by a combination of a 2D optical lattice and an harmonic magnetic or optical dipole trap \cite{Wang09}. The lattice parameters along its principle axes $x$ and $y$ can be tuned so as to give strong hopping along $x$ and weak hopping along $y$, so as to generate a pair of weakly coupled double-well potentials. Another option would be to employ double-well potentials with two spin states in each well. In this case one can tune the optical or magnetic coupling between the spin states in each well to be much larger or much smaller than the inter-well hopping rate. A variation on this configuration would be to utilize two spatial modes in each well instead of two spin states.
 
In terms of preparation, two-mode BECs have already been prepared in coherent states \cite{Schumm05,Lucke11}, squeezed states \cite{Esteve08,Gross10,Riedel10,Berrada13} and Fock states \cite{Lucke11} for atom interferometry purposes.  Moreover, the ability to rotate both preparation arbitrarily on the Bloch sphere has also been demonstrated in the lab. Similar techniques can be used to separately manipulate the two dimers. For example, it is possible to separate the two wells in the double-well, two spin states configuration and use the optical coupling between the internal spin states to engineer the desired entropy gradient, eventually re-coupling the wells. Alternatively one can begin with an appropriate four-mode coherent state and reduce the one particle coherence of one dimer by decreasing its internal coupling so as to enable an interaction-induced one-axis twist \cite{Kitagawa93,Gross10,Riedel10} followed by $\pi/2$ rotation about $S_{+,y}$ to transform the number ($S_{+,z}$) squeezed state into the required odd-even ($S_{+,x}$) squeezed state.

Finally, we comment on the required Fock state in the low-entropy dimer. Experimentally, twin-Fock states with equal mode populations are easier to prepare.  There is no restriction on using such states as long as they still lie within the linear excitation regime of the pertinent dimer, i.e. as long as $u_\Omega\ll 2$. In particular, all results shown here for $u_\Omega=0.05$ apply also to a 'coherent plus twin-Fock' preparation.
 
\section{Conclusion} 
We predict a {\em quantum} oscillation mode between weakly-coupled Bose-Hubbard dimers prepared with the same particle number and Josephson excitation energy, but with different one-particle coherence. Assuming that one dimer prepared in a coherent state, can be coupled to another dimer prepared in a (rotated) Fock state, which is approximately a dimer energy eigenstate, we expect an oscillation of the one particle purity between the dimers at the joson frequency $\tilde\omega_\nu$. While this phenomenon goes beyond the classical mean-field approximation, it is essentially {\em semiclassical}, resulting from the interplay of the two fast Bogoliubov modes of the BHDD. In the presence of interactions, the beat frequency between these two mode is rescaled by the effective joson-joson attraction $U_\nu$. Moreover, interactions serve as an effective heat-bath for the entropy oscillations, eventually leading to their equilibration and to the loss of single-particle coherence.

{\acknowledgments}
We acknowledge valuable discussions with James Anglin and Yoni Dubi.  This research was supported by the Israel Science Foundation (ISF grant No. 346/11) and United States-Israel Binational Science Foundation (BSF grant No. 2008141).


\end{document}